\documentclass[]{spie}  %>>> use for US letter paper
\usepackage[]{graphicx,latexsym,color}

\newcommand{\etal}{et al.}

\newcommand{\eg}{e.g.}
\newcommand{\ie}{i.e.}

\newcommand{\micron}{\mbox{$\mu{\rm m}$}}

\newcommand{\Mjup}{\hbox{M$_{\rm Jup}$}}
\newcommand{\meth}{{\hbox{CH$_4$}}}   % CH4

\newcommand{\Teff}{\mbox{$T_{\rm eff}$}}

\newcommand{\arcsec}{\mbox{$^{\prime \prime}$}}

\def\lesssim{\mathrel{\hbox{\rlap{\hbox{%
 \lower4pt\hbox{$\sim$}}}\hbox{$<$}}}}
\def\gtrsim{\mathrel{\hbox{\rlap{\hbox{%
 \lower4pt\hbox{$\sim$}}}\hbox{$>$}}}}

\newcommand\aj{AJ}% 
\newcommand\apj{ApJ}% 
\newcommand\apjl{ApJ}% 
\newcommand\apjs{ApJS}% 
\newcommand\aap{A\&A}% 
\newcommand\mnras{MNRAS}% 
\newcommand\pasp{PASP}% 

\title{The Gemini NICI Planet-Finding Campaign}

\author{
Michael C. Liu,\supit{1}
Zahed Wahhaj,\supit{1}
Beth A. Biller,\supit{1}
Eric L. Nielsen,\supit{2}
Mark Chun,\supit{3}
Laird M. Close,\supit{2}
Christ Ftaclas,\supit{1}
Markus Hartung,\supit{4}
Thomas L. Hayward,\supit{4}
Fraser Clarke,\supit{5}
I. Neill Reid,\supit{6}
Evgenya L. Shkolnik,\supit{7}
Matthias Tecza,\supit{5}
Niranjan Thatte,\supit{5}
Silvia Alencar,\supit{8}
Pawel Artymowicz,\supit{9}
Alan Boss,\supit{6}
Adam Burrows,\supit{10}
Elisabethe de Gouveia Dal Pino,\supit{11}
Jane Gregorio-Hetem,\supit{11}
Shigeru Ida,\supit{12}
Marc J. Kuchner,\supit{13}
Douglas Lin,\supit{14}
Douglas Toomey\supit{15}
\skiplinehalf
{\small
  \supit{1}{Institute for Astronomy, University of Hawaii, 2680
    Woodlawn Drive, Honolulu, HI 96822}\\
  \supit{2}{Steward Observatory, University of Arizona, 933 North Cherry Avenue, Tucson, AZ 85721}\\
  \supit{3}{Institute for Astronomy, 640 North A‘ohoku Place, \#209, Hilo, Hawaii 96720-2700 USA}\\
  \supit{4}{Gemini Observatory, Southern Operations Center, c/o AURA, Casilla 603, La Serena, Chile}\\
  \supit{5}{Department of Astronomy, University of Oxford, DWB, Keble Road, Oxford OX1 3RH, U.K.}\\
  \supit{6}{Space Telescope Science Institute, 3700 San Martin Drive, Baltimore, MD 21218}\\
  \supit{7}{Department of Terrestrial Magnetism, Carnegie
    Institution of Washington, 5241 Broad Branch Road, NW, Washington,
    DC 20015}\\
  \supit{8}{Universidade Federal de Minas Gerais}\\
  \supit{9}{University of Toronto at Scarborough, 1265 Military Trail, Toronto, Ontario M1C 1A4, Canada}\\
  \supit{10}{Department of Astrophysical Sciences, Peyton Hall, Princeton University, Princeton, NJ 08544}\\
  \supit{11}{Universidade de Sao Paulo, IAG/USP, Departamento de Astronomia, Rua do Matao, 1226, 
    05508-900, Sao Paulo, SP, Brazil} \\
  \supit{12}{Tokyo Institute of Technology}\\
  \supit{13}{NASA Goddard Space Flight Center, Exoplanets and Stellar Astrophysics Laboratory, Greenbelt, MD 20771}\\
  \supit{14}{UC Santa Cruz}\\
  \supit{15}{Mauna Kea Infrared, LLC, 21 Pookela St., Hilo, HI 96720}
}  % \small   
}

\authorinfo{Email: {\tt mliu@ifa.hawaii.edu}}
\begin{document} 
  \maketitle 

%%%%%%%%%%%%%%%%%%%%%%%%%%%%%%%%%%%%%%%%%%%%%%%%%%%%%%%%%%%%% 
\begin{abstract}
  Our team is carrying out a multi-year observing program to directly
  image and characterize young extrasolar planets using the
  Near-Infrared Coronagraphic Imager (NICI) on the Gemini-South
  8.1-meter telescope. NICI is the first instrument on a large telescope
  designed from the outset for high-contrast imaging, comprising a
  high-performance curvature adaptive optics (AO) system with a
  simultaneous dual-channel coronagraphic imager. Combined with
  state-of-the-art AO observing methods and data processing, NICI
  typically achieves $\approx$2 magnitudes better contrast compared to
  previous ground-based or space-based planet-finding efforts, at
  separations inside of $\approx$2\arcsec. In preparation for the
  Campaign, we carried out efforts to identify previously unrecognized
  young stars as targets, to develop a rigorous quantitative method for
  constructing our observing strategy, and to optimize the combination
  of angular differential imaging and spectral differential imaging. The
  Planet-Finding Campaign is in its second year, with first-epoch
  imaging of 174~stars already obtained out of a total sample of
  300~stars. We describe the Campaign's goals, design, target selection,
  implementation, on-sky performance, and preliminary results. The NICI
  Planet-Finding Campaign represents the largest and most sensitive
  imaging survey to date for massive ($\gtrsim$1~\Mjup) planets around
  other stars. Upon completion, the Campaign will establish the best
  measurements to date on the properties of young gas-giant planets at
  $\gtrsim$5--10~AU separations. Finally, Campaign discoveries will be
  well-suited to long-term orbital monitoring and detailed
  spectrophotometric followup with next-generation planet-finding
  instruments.
\end{abstract}

%>>>> Include a list of keywords after the abstract 


%%%%%%%%%%%%%%%%%%%%%%%%%%%%%%%%%%%%%%%%%%%%%%%%%%%%%%%%%%%%%
\section{INTRODUCTION}
\label{sec:intro}  % \label{} allows reference to this section

Radial velocity (RV) and transit detections of extrasolar planets have
been a watershed for observational studies of planet formation,
compiling a sample of planets large enough ($\gtrsim$400 to date) for
statistical studies. However, these discoveries provide an incomplete
picture of the extrasolar planet population: most RV planets are
detected only indirectly and with the $\sin i$ ambiguity in their masses,
and transiting planets are mostly restricted to very small orbital
separations. Direct imaging of exoplanets can measure colors,
luminosities and spectra, thereby providing temperatures and
compositions. Furthermore, since RV and transit studies are confined to
the inner regions of other solar systems ($<$6 AU for 15-yr survey), we
know very little about the planetary constituents in the outer regions
of other solar systems, where gas-giant planets are born.

The discovery of extrasolar planets by direct imaging
(\citenum{marois08-hr8799bcd, 2008Sci...322.1345K,2009A&A...493L..21L,
  2008ApJ...689L.153L}) has opened the door to a whole new realm of
observational study. Analogous to the growth of RV and transit studies,
the next steps in the field of direct imaging will be to move from
individual ``headline'' discoveries to well-defined, well-studied
samples to glean the properties of the whole population. Moreover,
detailed photometric and spectroscopic analysis of new exoplanet
discoveries will allow us to dissect the atmospheric properties and
thermal evolution of these objects (\eg, \citenum{2010ApJ...716..417H,
  2010arXiv1006.3070L, bowler10-hr8799b}).

%%%%%%%%%%%%%%%%%%%%%%%%%%%%%%%%%%%%%%%%%%%%%%%%%%%%%%%%%%%%%
\section{The Instrument}

In principle, the largest (8--10~meter) ground-based telescopes equipped
with adaptive optics (AO) could be effective for direct imaging of
planets, as these platforms achieve the highest possible angular
resolution in the near-IR with a filled aperture telescope. However,
traditional AO imaging is hampered by the time-variable nature of the
point spread function (PSF) and the presense of quasi-static point-like
speckles in the images. Thus while AO greatly enhances the contrast
(\ie, the ability to detect faint sources next to bright ones), its
imperfect correction is a severe limitation to push to planetary masses
and separations. This challenge can be overcome with specialized
instrumentation.

\begin{figure}[t]
\includegraphics[width=4.2in,angle=0]{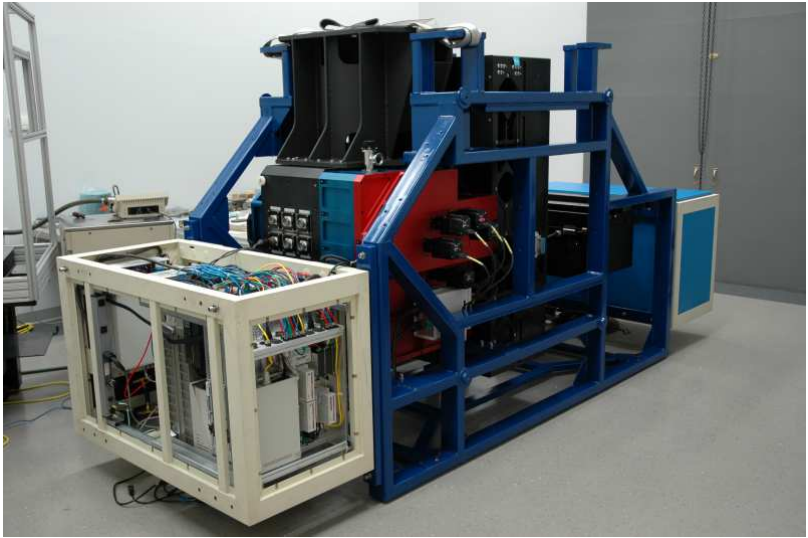}
\hskip 0.2in
\raise -0.1in 
\vbox{\includegraphics[width=2.2in,angle=0]{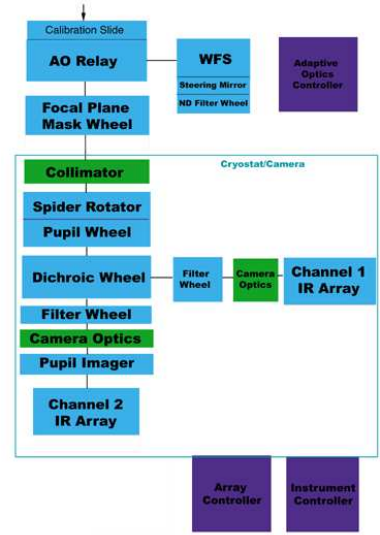}}
%\vbox{\includegraphics[width=2.2in,angle=0]{NICI_blockdiagram.v2.ps}}
%\vbox{\includegraphics[width=2.2in,angle=0]{NICI_blockdiagram_small.epsf}}
%\vbox{\includegraphics[width=2in,angle=0]{NICI_blockdiagram.ps}}
\vskip 2ex
\caption{\small \em {\bf Left:} NICI at Mauna Kea Infrared in Hilo,
  Hawaii, prior to acceptance testing. {\bf Right:} Block diagram of
  NICI optical configuration, including the 85-element curvature AO
  system, the focal \& pupil plane mask mechanisms, and the two IR
  imaging channels. NICI is the first imager for a 8-10~meter telescope
  designed expressly for exoplanet imaging and is now in routine
  operation at the Gemini-South 8.1-meter Telescope. \label{fig:nici}}
\vskip -1ex
\end{figure}

The Near-Infrared Coronagraphic Imager (NICI) is a powerful AO
instrument tailored to direct detection of extrasolar planets through
high contrast imaging ({\bf Figure~\ref{fig:nici}}). It is the first
instrument on an 8-10~meter telescope designed expressly for such work.
NICI was built by Mauna Kea Infrared in Hilo, Hawaii (PI Doug Toomey)
and funded by NASA. The instrument is now fully operational at the
Gemini-South 8.1-meter telescope\cite{2008SPIE.7015E..49C}. NICI
combines a suite of capabilities to achieve high contrast imaging:
(1)~an efficient natural guide star curvature AO system built by the
University of Hawaii, (2) spectral differential imaging (SDI); (3)
angular differential imaging (ADI); and (4) Lyot-style coronography.
While these techniques have been used in previous instruments, NICI is
the first to bring all of them together into a single instrument.

NICI was designed as a complete end-to-end system for high-contrast
imaging, minimizing both the wavefront phase distortions from the
atmosphere, telescope and instrument as well as the internal
instrumental scatter. NICI first creates a high Strehl image
($\approx$30--45\% at $H$-band) with its own internal low-scatter
85-element curvature AO system. Its wavefront sensor is tailored to the
range of natural guide star brightnesses needed for the Planet-Finding
Campaign ($V\lesssim14$~mag). Unlike most other AO systems to date, the
corrected AO beam is reflected into the science channel and the first
transmissive element is the focal-plane mask. The (warm) focal plane
mechanism offers several choices of circular translucent masks, all with
flat-topped gaussian transmission profiles and central attenuations of
order 0.5\%. The masks effectively boost the dynamic range of the
detector, allowing us to accurately determine the position and flux of
the central star relative to any faint companion candidates. NICI's
internal (cryogenic) pupil mechanism allows for several hard-edged
stops, which help to remove PSF artifacts associated with the edges of
the Gemini-South secondary mirror. Following the focal-plane mask and
pupil stop, the beam is divided with a 50/50 beam-splitter into two
imaging channels that are read out simulataneously.

\begin{figure}
%\vskip -0.5in
\centerline{\includegraphics[width=5in,angle=0]{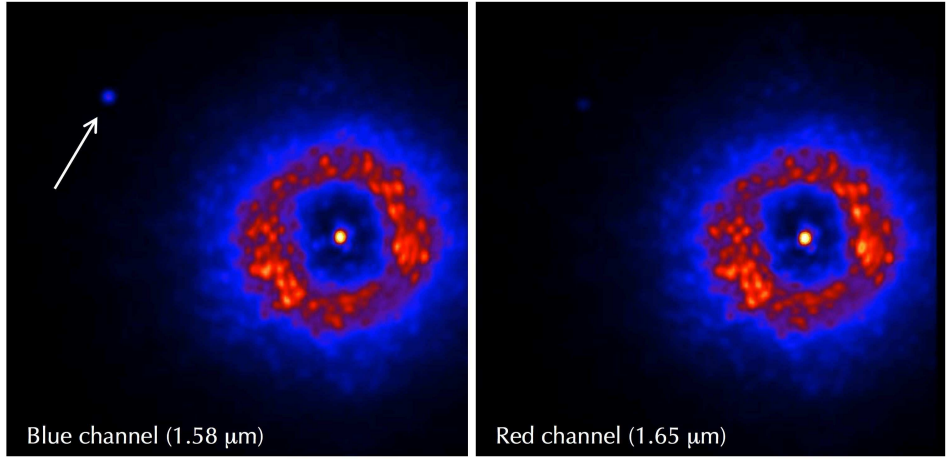}}
%%\centerline{\includegraphics[width=5in,angle=0]{dual+adi+sdi-demo-200pix-h254bcx19.eps}}
\vskip 2ex
\caption{\small \em An example of NICI two-channel imaging data from a
  1-minute exposure of a bright star. 
  %The left image shows the blue
  %imaging channel with the 1.578~\micron\ methane-off filter, and the
  %right image shows the red channel with the 1.652~\micron\ methane-on
  %filter. 
  The FOV of each image is 3.6\arcsec\ on a side. The science target is
  centered on the translucent focal plane mask, and the highly speckled
  nature of the PSF halo is seen in the images from both channels. An
  artificial methane-bearing companion has been inserted into the data
  in the upper left of each image; it is brighter in the left image than
  the right one due to the presence of methane absorption in the red
  channel filter. \label{fig:dualchannel}}
%\vskip -0.4in
\end{figure}

For ages of $\lesssim$100~Myr, ultracool ($\Teff<1300$~K) objects with
near-IR photospheric \meth\ absorption correspond to masses of
$\lesssim$12~\Mjup\ according to evolutionary models (\eg,
\citenum{1997ApJ...491..856B}), and hence \meth\ absorption is expected
to be a characteristic signature of young planets. IR imaging of young
planets in and out of this absorption band will produce a unique
photometric signature, strong emission in the blue band and little
emission in the red one, that can be distinguished from the bright
(methane-free) glare of the parent star \cite{1999PASP..111..587R}.
The SDI approach was first attempted with the Trident camera on CFHT
\cite{2003IAUS..211..275M} and has also been used for the SDI cameras on
the VLT and MMT \cite{2004SPIE.5492..970L, 2004SPIE.5490..389B}. Similar
to these instruments, NICI provides time simultaneous methane-band
imaging in order to counteract the time-variable AO PSF. NICI implements
SDI through dual-channel imaging design, with each channel being an
independent optical channel with its own $1024\times1024$ ALADDIN InSb
detector ({\bf Figure~\ref{fig:dualchannel}}).
The spectral properties of the $H$-band on+off methane filters in NICI
were custom designed to maximize the combination of sensitivity and
accurate SDI subtraction, based on an end-to-end simulation of the
expected imaging performance (see \citenum{2008SPIE.7015E..49C}).  The
resulting filters are 4\% wide, with central wavelengths of
1.578~\micron\ (off-methane) and 1.652~\micron\ (on-methane).

NICI can also employ ADI (a.k.a. roll subtraction;
\citenum{2004Sci...305.1442L, 2006ApJ...641..556M}) to distinguish
between long-lived telescope+instrument speckles and faint astronomical
objects, thereby removing the PSF and achieving higher contrast.
Altogether, NICI offers a number of imaging options, through the use of
ADI, SDI, or both; this versatility makes it novel compared to other
previous AO instruments. For instance, dual-channel data obtained with
ADI+SDI mode can also be summed, instead of differenced as in the SDI
processing, to yield a pseudo-broadband ADI dataset. Likewise, the
instrument can be used in single-channel ADI-only mode, where an
internal mirror is used to send all the light to one detector, thereby
maximizing the throughput. {\em NICI's versatile imaging configurations
  make it sensitive to very faint companions, both with and without
  photospheric methane absorption.}

%%%%%%%%%%%%%%%%%%%%%%%%%%%%%%%%%%%%%%%%%%%%%%%%%%%%%%%%%%%%%
\section{Goals and Strategy}

To take full advantage of the powerful capabilities of NICI, we are
leading a three-year guaranteed-time campaign at the Gemini-South
8.1-meter Telescope dedicated to finding and characterizing planets by
direct imaging.
The Campaign is designed to address three key questions in the study of
gas-giant extrasolar planets:

\begin{enumerate}

\item {\bf \em What is the frequency of outer ($>$5--10~AU) massive
    planets around other stars?} Determining the incidence and
  properties of outer planetary companions will allow us to develop a
  complete picture of exoplanetary configurations.
To this end, one major goal of the NICI Campaign is to probe the mass
and separation distribution ($dN/dM/da$) of planets at distances as
close as $\gtrsim$5--10~AU, as inferred from the complete set of NICI
detections (discoveries) and non-detections.
This distribution may have profound consequences for assessing the
dominant formation mechanism of gas-giant planets.

% - - - - - - - - - - - - - - - - - - - - - - - - - - - - - - - %

%\item 
\item {\bf \em What is the dependence of planet frequency on the stellar
    host mass}? The frequency of giant planets in the outer regions of
  low-mass stars (M~dwarfs) is another key discriminant between the two
  competing theories of giant planet formation, namely core accretion
  and disk instability. By design, the Campaign is searching for planets
  around young stars over a wide range of masses, from spectral type B7
  to M6. This is feasible thanks to the sensitivity of NICI's
  curvature-based AO system to optically faint stars. RV surveys find
  few massive planets in the inner $\lesssim$1~AU regions around
  low-mass stars (\eg, \citenum{2010arXiv1005.3084J,
    2007ApJ...665..785J}); the NICI Campaign will be a complementary
  study of the outer regions around these objects.

% - - - - - - - - - - - - - - - - - - - - - - - - - - - - - - - %

%\item 
\item {\bf \em What are the spectrophotometric properties of young
    extrasolar planets}? Follow-up multi-band photometry and
  spectroscopy of directly imaged planets will test theoretical models,
  which are far from mature. The discovery space is large and
  unexplored. Cooling models may be incorrect or missing key opacity
  sources. Indeed, one of the early surprises from RV discoveries was
  the diversity of exoplanet orbits. Whether this diversity extends to
  their spectral energy distributions (SEDs) is an important open
  question. Initial studies of the SEDs of the HR~8799 planets point to
  unusually cloudy, non-equilibrium photospheres compared to field brown
  dwarfs, suggesting extreme physical properties in ultracool
  atmospheres at young ages (\citenum{marois08-hr8799bcd,
    2010ApJ...716..417H, bowler10-hr8799b}). However, many more systems
  are needed for study.

\end{enumerate}

\begin{figure}[t]
%\vskip -0.1in
%\hskip -0.6in
%\includegraphics[width=2.45in,angle=0]{nielsen-nici_target43_msplot01_log.ps}
%\hskip -0.15in
%\includegraphics[width=2.45in,angle=0]{nielsen-trends_age_nici03.ps}
%\hskip -0.15in
%\includegraphics[width=2.45in,angle=0]{nielsen-trends_dist_nici02.ps}
%\hskip -0.4in
\includegraphics[width=3in,angle=0]{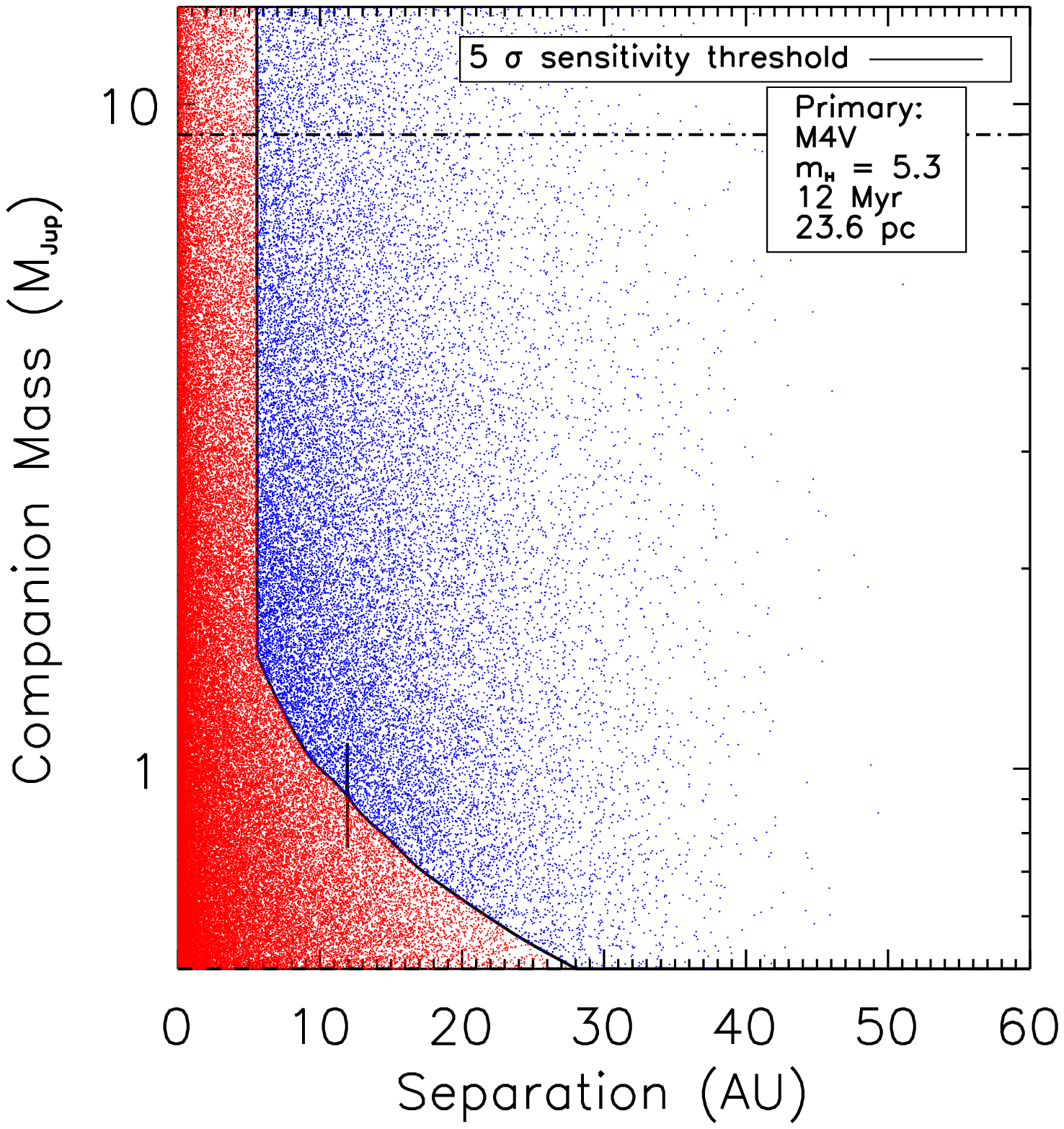}
\hskip 0.2in
\includegraphics[width=3in,angle=0]{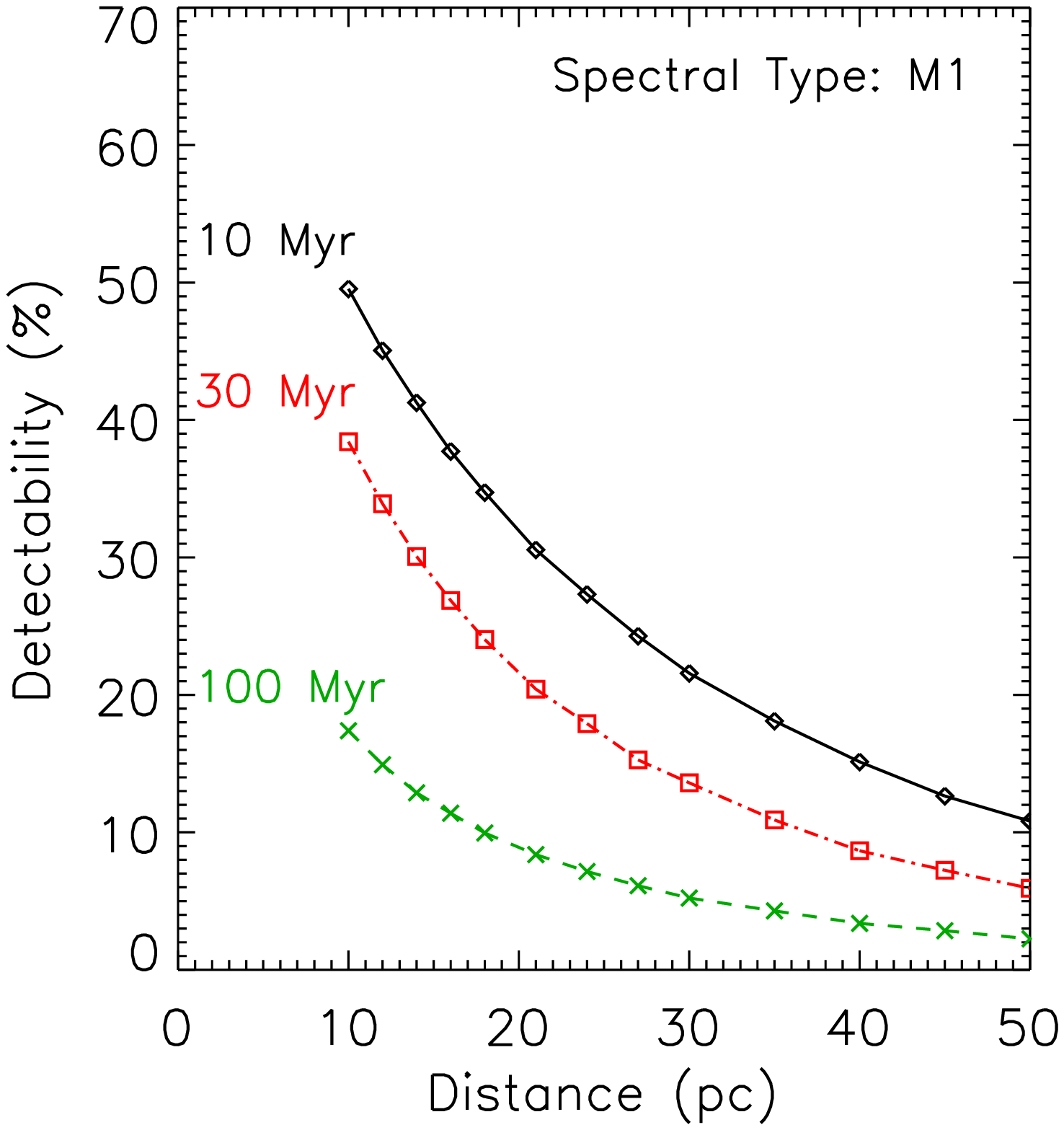}
%\vskip -0.2in
\caption{\small \em Illustrations of our Monte Carlo simulations for
  Campaign target selection and prioritization, based on the methodology
  of Nielsen \etal\ (\eg, \citenum{2008ApJ...674..466N}). {\bf Left:} A set of
  100,000 simulated planets are shown, with blue points marking planets
  above the detection threshold, and red points marking planets not
  detected by NICI. For this particular star, 25\% of simulated planets can be
  detected, \ie, Detectability~=~25\%. We carried out these detailed
  simulations for all stars on our initial target list. 
  %{\bf Middle:}
  %The fraction of planets detected as a function of target star age, for
  %a semi-major axis distribution $dN/da \propto a^{-0.5}$ from
  %0.03--30~AU, and 3~choices of host star spectral type. {Younger and
  %  lower-mass stars are more favorable for detection}. 
  {\bf Right:} The fraction of planets detected around an M1~primary as
  a function of target star distance, showing the trade-off with target
  age for an input planet population with a semi-major axis distribution
  $dN/da \propto a^{-0.5}$ from 0.03--30~AU. The nearest, youngest stars
  are favored, though more nearby, slightly older  stars can be better
    targets than younger stars farther away. \label{fig:simulations}}
\end{figure}

% %----------------------------------------------------------------------%
% %----------------------------------------------------------------------%
% %

% \section{Observing Campaign Description and Status}

To achieve these goals, the NICI Campaign is mostly targeting nearby
young ($\lesssim$300~Myr) stars, where planets are expected to be hot
enough and luminous enough for direct detection at near-IR wavelengths.
One added virtue of NICI is that it is deployed at Gemini-South. While
numerous young stars have been recognized in the last 5--10~years all
over the sky, the most promising of the currently known moving groups
reside in the southern hemisphere (\eg, \citenum{1995MNRAS.273..559J,
  2001ApJ...562L..87Z, 2008hsf2.book..757T}).

We also carried out complementary efforts to identify previously
unrecognized young stars as targets prior to the start of the Campaign,
focusing on low-mass stars (M~dwarfs) within
25~pc\cite{2009ApJ...699..649S}.
The current young star census is mostly restricted to higher-mass
(AFGK-type) stars and contains few M dwarfs. This paucity is striking,
especially considering that M dwarfs dominate the stellar mass function
by number: M~dwarfs comprise $\approx$70\% of a volume-limited census
(\eg, \citenum{2008AJ....136.1290R}).
To find this ``missing'' population, we used X-ray activity+color
selection to identify candidates and high-resolution optical
spectroscopy to refine their age estimates, through gravity-sensitive
indices, strong H$\alpha$, lithium, and $UVW$ space velocities. The vast
majority of our M~dwarfs are not in any previously published young star
sample, illustrating the novelty of our search.

To select and prioritize targets, prior to the start of the Campaign we
employed Monte Carlo simulations to evaluate the science return of
different approaches: deeper vs. shallower exposures; more vs.\ fewer
targets; younger, more distant targets vs. older, closer targets, etc.
Our approach is based on the methods developed by Nielsen \etal\ (\eg,
\citenum{2008ApJ...674..466N}). The ranking of targets is done by
simulating a large number of planets (100,000) around each star, with
planets drawn from mass, semi-major axis, and eccentricity distributions
consistent with the known RV planet population and null results from
previous direct imaging surveys\cite{2010ApJ...717..878N}.
The brightness, age and distance of the host star establish the apparent
magnitudes, flux ratios, and projected separations of the simulated
planets.
The simulated planets were then compared with the expected NICI
companion sensitivity to determine what fraction would be detected ({\bf
  Figure~\ref{fig:simulations}}).
This approach
allows us to understand the trade-offs of the relevant factors: age,
distance, host star luminosity, AO performance, exposure times, and
overall sample size. The result is that the best stars can be identified
for deep NICI imaging, out of thousands of possible targets.
In short, the NICI Campaign has been designed to maximize the likelihood
of detecting planets; as a natural consequence, this also ensures that
even a null result would have profound scientific impact, strongly
constraining the possible populations of long-period giant extrasolar
planets.

The final Campaign target list is composed primarily of stars with ages
of $\lesssim$300~Myr and distances of $\lesssim$70~pc. It does include
stars with older ages or larger distances that are promising targets,
especially if they have ancillary evidence for being hosts of planetary
systems (\eg, the presence of circumstellar debris disks). Our
simulation effort showed that the ``best'' list of stars depends to some
degree on the assumed input planet population, especially the adopted
outer orbital separation and its dependence (or lack thereof) on stellar
host mass. This was not surprising, though the full extent of the effect
is perhaps unappreciated in previous such simulations --- the choices
that go into the modeling inevitably sway the outcome. In the end, we
synthesized the results from simulations with different assumptions,
ensuring the broad range of spectral types needed to study the
dependence of planet frequency on stellar host mass (Campaign goal \#2).
The final sample is split roughly equally between high-mass stars (AF
spectral types), solar-type stars (GK types), and low-mass stars (M
type).

%%%%%%%%%%%%%%%%%%%%%%%%%%%%%%%%%%%%%%%%%%%%%%%%%%%%%%%%%%%%%
\section{Status}

% --- Timeline ---%

The Campaign began science observations in December 2008, with monthly
observing runs executed in fixed blocks of several nights during bright
time. Observations are executed on-site by Gemini staff, with real-time
remote support by Campaign team members via Polycom.
Most NICI observations have been carried out during the Chilean summer,
from November to April when seeing conditions are favorable for AO
imaging.
First-epoch observations for 178 stars have been obtained so far, more
than half the Campaign goal of 300 stars.

%--- How observing done ---%

Thanks to the extensive on-sky characterization during the commissioning
phase and first year of the Campaign, we have developed a stable set of
observing protocols that are now used for almost all Campaign
observations\cite{2008SPIE.7015E.184B}, ensuring homogenous datasets
that are directly amenable to prompt pipeline processing and common
science analysis.
Most targets are observed contemporaneously with two instrument
configurations: (1)~a dual-channel ADI+SDI mode (``ASDI'') with the 4\%
1.6~\micron\ methane on+off filters and (2)~a single-channel ADI-only
mode with the regular $H$-band filter. This ``hybrid'' scheme provides
the greatest sensitivity over a range of separations. The ASDI mode
delivers the highest contrasts in the inner $\lesssim$1.0--1.5\arcsec\
where speckle noise dominates, while the ADI-only mode offers the
greatest sensitivity to faint companions at larger separations.

\begin{figure}
\begin{center}
\includegraphics[width=3.5in,angle=0]{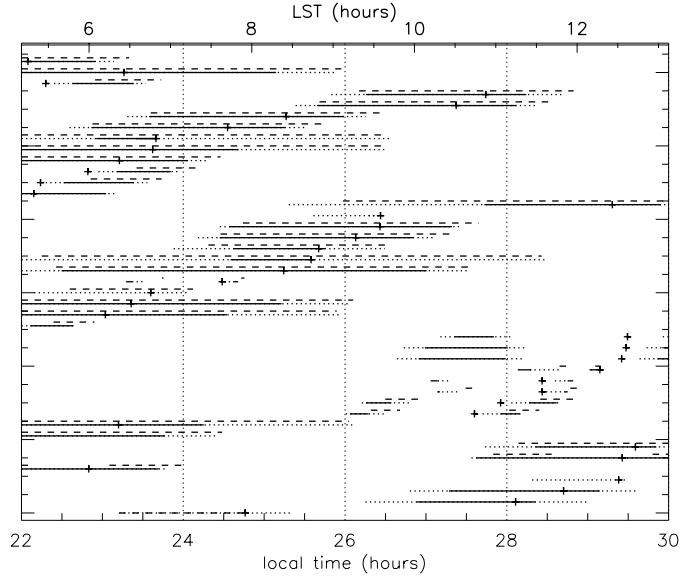}
\end{center}
\caption{\small \em An example of the observing windows for a typical
  NICI Campaign night. Each set of dotted+dashed+solid lines represents
  the allowable start time for one science target, where the sky
  rotation is matched to our ADI requirements on the instantaneous
  rotation rate and the total amount of rotation for each dataset. (The
  different line styles corresponded to slightly different
  calculations.) The “+” sign marks the transit for each target. Each
  observing night is planned to optimize the target priorities, given
  the feasible observing windows. \label{fig:schedule}}
%\vskip -0.4in
\end{figure}

Each night, we design an observing plan to carefully control the amount
of sky rotation for each target ({\bf Figure~\ref{fig:schedule}}). Too
much instantaneous rotation during a single ADI exposure will lead to
too much blurring and thus loss of point-source sensitivity at larger
separations. Too little total rotation over the entire observing
sequence for a target will make it difficult to construct an appropriate
PSF for ADI data reduction and lead to self-subtraction at smaller
separations. Fundamentally, the duration of an observing window is a
function of target declination. For objects that transit close to
overhead, these windows can be very brief, sometimes only 10-20 minutes
long. For more northerly or southerly targets, the observing windows are
longer, up to several hours in duration, and therefore easier to
schedule. Given a prioritized list of science targets, we assemble a
custom schedule for each night, fitting together the differently sized
observing windows for all the targets. This ensures that objects with
difficult (short) observing windows will be observed with an optimal
rate and total amount of sky rotation.

%--- Data processing ---%

\begin{figure}[t]
\hskip -0.6in 
\includegraphics[width=3.2in,angle=90]{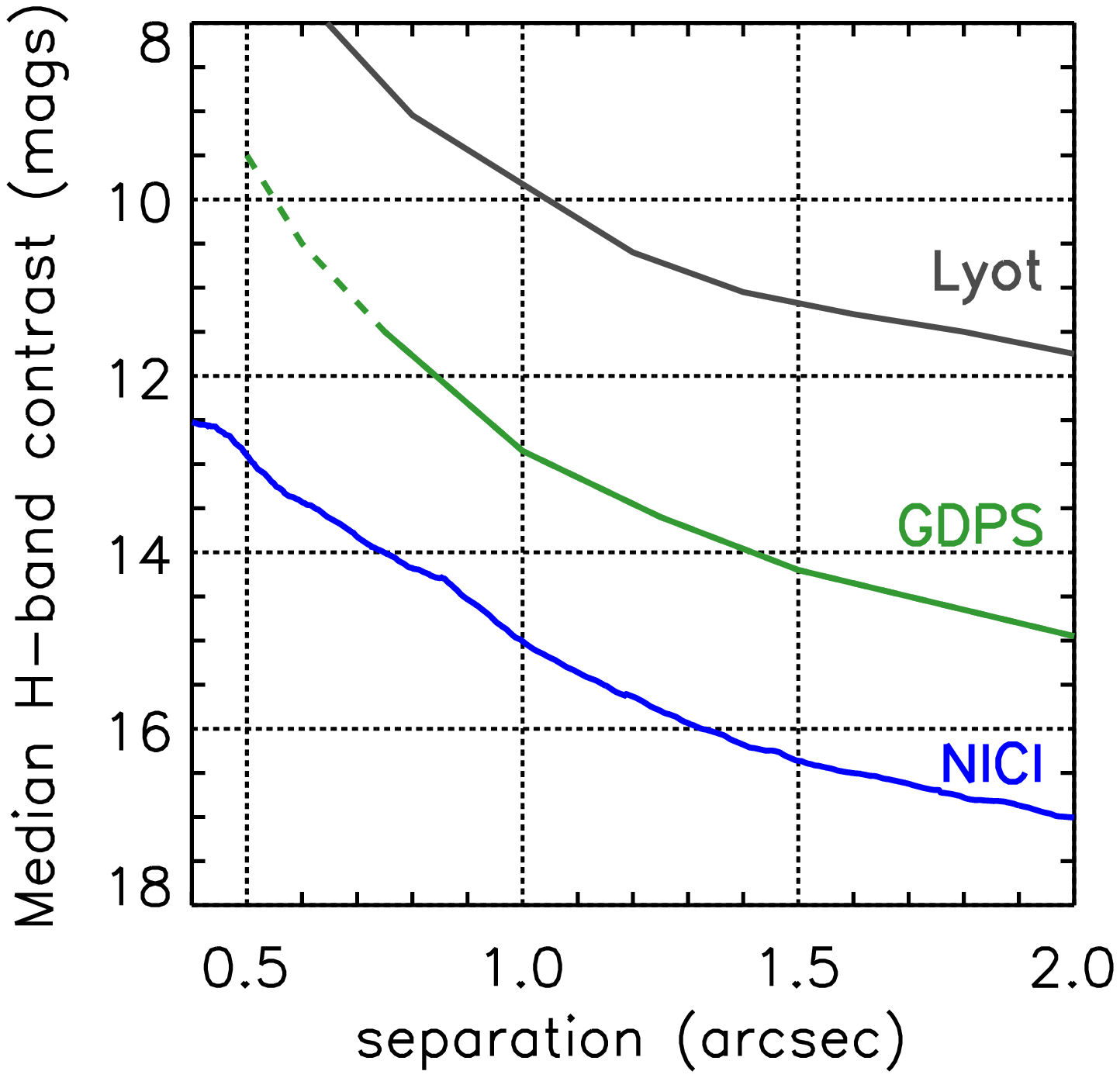}
\hskip -1.1in
\includegraphics[width=3in,angle=90]{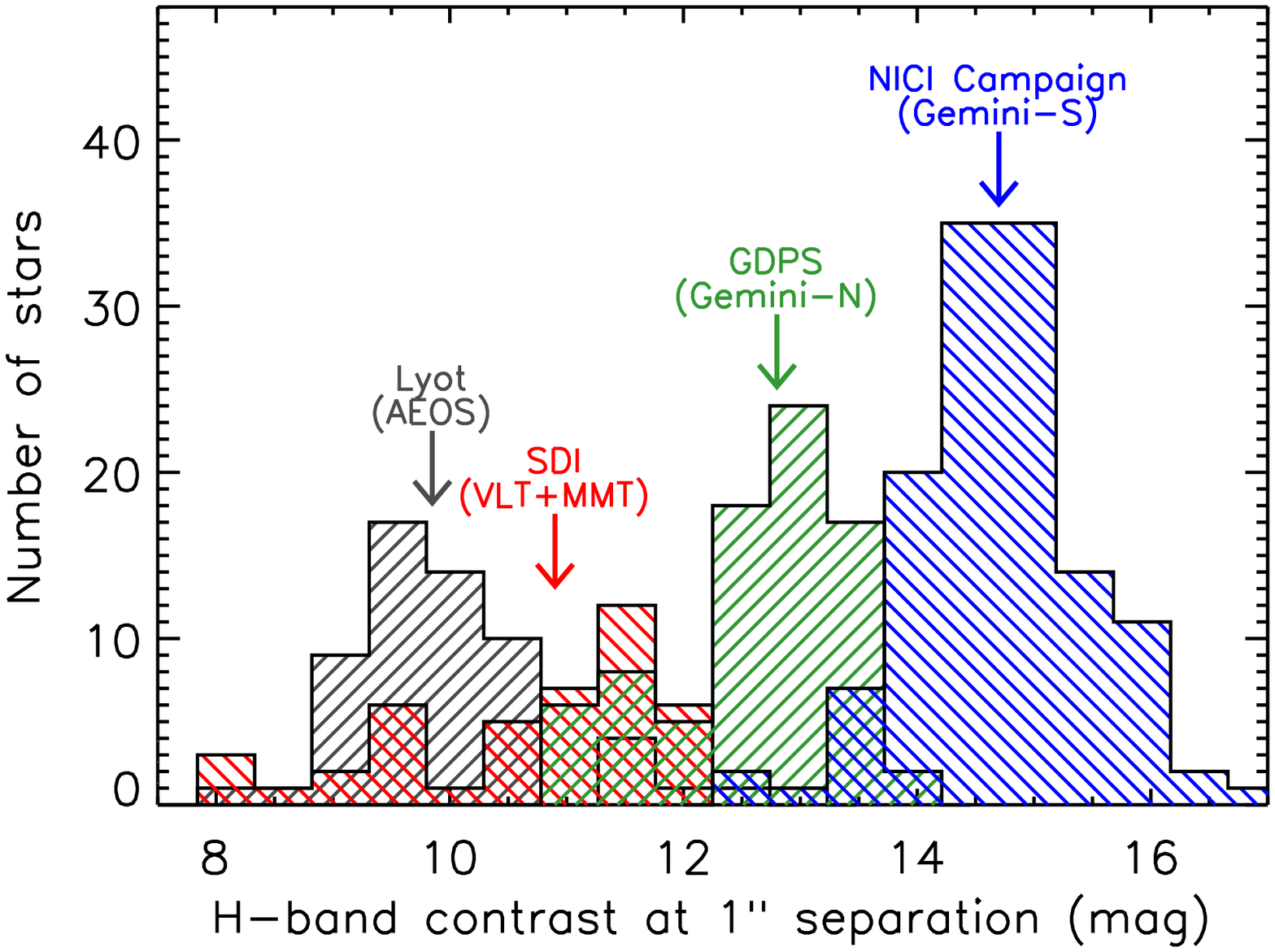}
%\includegraphics[width=3.5in,angle=0]{candidate_masses_hist_closest.ps}
%\vskip -2ex
\caption{\small \em {\bf Left:} The \underline{median} $H$-band contrast
  curve for point-source detection from the NICI Campaign so far
  (typical on-source integration time of 45~min), compared to the Gemini
  Deep Planet Survey conducted at Gemini-North\cite{2007ApJ...670.1367L}
  and the Lyot Project at the AEOS Telescope\cite{2010ApJ...716.1551L}.
  The NICI Campaign curve merges the dual-channel ASDI data taken with
  the 4\% methane filters and single-channel ADI datasets taken with the
  standard $H$-band filter. For GDPS, the data were obtained with a 6\%
  methane-off (1.58~\micron) filter. Also, many of the GDPS stars were
  saturated in the innermost regions, hence the dashed line. {\bf
    Right:} Comparison of the NICI Campaign so far with the
  aforementioned surveys and also the VLT+MMT/SDI
  survey\cite{2007ApJS..173..143B}. The plotted histogram shows the
  point-source detectability at 1\arcsec\ separation for every star
  observed by these surveys. The NICI Campaign is not yet completed
  (only about 1/2 the sample has been observed), but is already the
  largest, most sensitive planet-imaging survey to
  date. \label{fig:contrast-histogram}}
%\vskip -0.4in
\end{figure}

Data processing and analysis occur immediately during and after each
run, providing the feedback on data quality and any potential
discoveries needed to plan the next block of observing.
with the combined angular+spectral differencing imaging (ASDI) mode and
the ADI-only mode. Individual images are scaled, registered, radially
aligned, and optimally differenced to subtract the PSF halo and speckles
from the bright star, thereby revealing any faint, close companions. The
pipeline is a fully working system, has already processed
$\approx$300~hours of Campaign observations, and has been vetted via
fake-companion injection+recovery experiments. The overall contrast
performance achieved by the NICI Campaign is a significant advance over
previous ground-based or space-based direct imaging surveys, by at least
a factor of $\approx$2~mag ({\bf Figure~\ref{fig:contrast-histogram}}).

%--- Second epoch followup ---%

\begin{figure}
\hskip 0.8in 
\includegraphics[width=3.3in,angle=90]{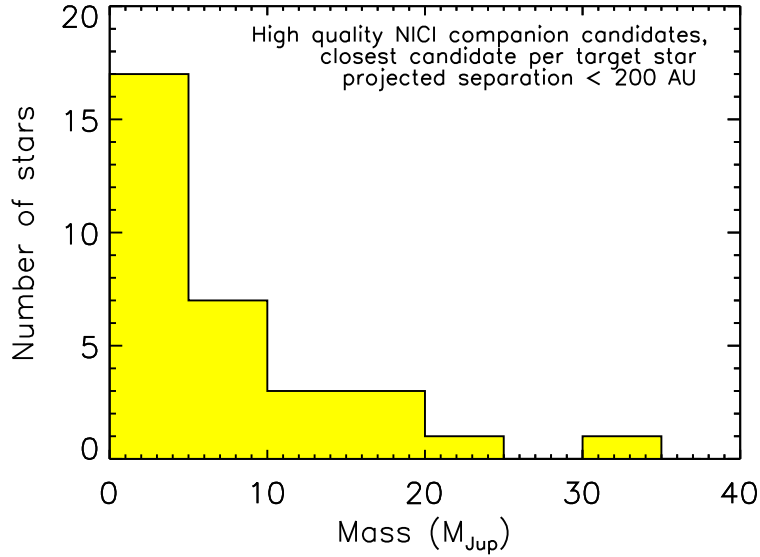}
%\includegraphics[width=4in,angle=0]{candidate_masses_hist_closest.ps}
%\vskip -2ex
\caption{\small \em Histogram of high-quality companion candidates found
  from Campaign first-epoch observations, with masses estimated from
  their absolute magnitudes, the estimated ages of their host stars, and
  hot-start evolutionary models\cite{1997ApJ...491..856B,
    2000ApJ...542..464C}. The plotted data comprise only stars at high
  galactic latitude with candidates at $<$200~AU separation and show
  only the closest candidate for each star. Most of these are expected
  to be uninteresting background stars, but some will be bona fide
  companions. Follow-up observations are currently
  underway.\label{fig:candidates}}
%\vskip -0.4in
\end{figure}

From first-epoch observations, the Campaign has identified many
high-quality substellar candidate companions ({\bf
  Figure~\ref{fig:candidates}}). ``High-quality'' here means that the
candidates pass multiple selection criteria: (1)~they are robust
detections, far above the false positive rate in the residual speckles
in processed data; (2)~they occur around stars with low stellar
backgrounds, as judged by an infrared galactic star count model (based
on \citenum{1980ApJS...44...73B, akraus2009-thesis}); and (3)~they have
a reasonable projected physical separation ($<$200~AU). The large number
of detections is not surprising, given the extreme depth of Campaign
observations.

\begin{figure}
\vskip 0.5in
\begin{center}
\includegraphics[width=5in,angle=0]{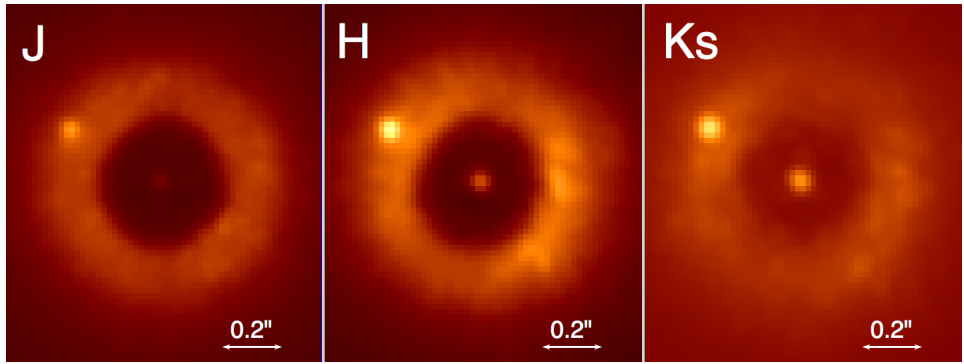}
\end{center}
\caption{\small \em NICI images of a newly discovered substellar
  companion to the young star PZ~Tel, a $\beta$~Pic moving group member
  (from \citenum{biller10-PZTelB}). The primary star resides at the
  center of the translucent focal plane mask. The companion is seen at a
  separation 0.36\arcsec\ (18~AU) in the 10 o'clock position. Two epochs
  of NICI imaging over 13 months has confirmed this as a common proper
  motion companion at very high confidence. A small amount of radial
  orbital motion is also detected, indicating a rather eccentric orbit
  for the companion ($e>0.6$). The estimated mass of PZ~Tel~B is
  36$\pm$6~\Mjup, based on its absolute magnitudes and the Lyon/DUSTY
  evolutionary models. This is one of the tightest substellar companions
  directly imaged to date, and thus is a promising system for long-term
  monitoring of orbital motion. \label{fig:pztel}}
%\vskip -0.4in
\end{figure}

\begin{figure}[h]
\vskip 0.2in
\hskip 0.2in 
\includegraphics[width=2.2in,angle=0]{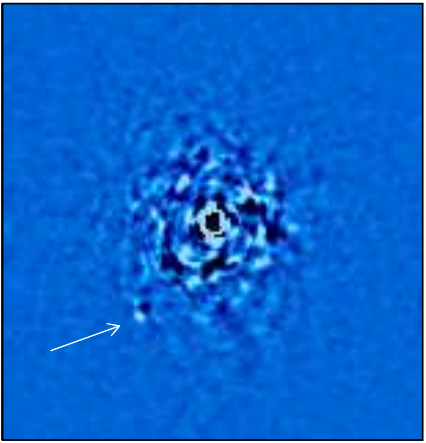}
\hskip 0.3in
\raise -0.4in
\vbox{\hsize=3in
\includegraphics[width=4in,angle=0]{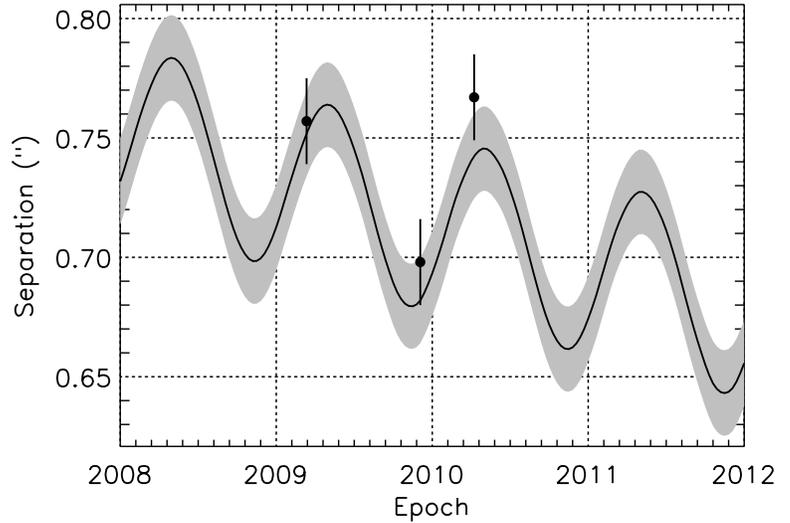}}
\vskip 2ex
\caption{\small \em {\bf Left:} A fully processed NICI 1.6~\micron\ ASDI
  dataset of the young star UY~Pic, a member of the 70~Myr AB~Dor moving
  group. The halo of the bright primary star has been removed by the
  Campaign pipeline. The FOV of this cutout is 2.7\arcsec\ on a side.
  (The full NICI FOV is 18.4\arcsec.) The arrow points
  to a very faint non-methanated companion, which has a
  positive/negative dipole appearance due to the SDI subtraction
  process. The companion has a projected separation of 0.8\arcsec\ and
  is 12.1 magnitudes fainter than the primary star. If physically
  associated, it would have had a mass of 7~\Mjup\ and a physical
  separation of 19~AU. {\bf Right:}~Three epochs of NICI astrometry of
  the faint companion, shown as the black points with error bars. The
  tilted sinusoidal curve shows the expected separation of the companion if it
  were a background object at infinite distance, given the known
  parallax and proper motion of the primary star and the first-epoch
  astrometric uncertainties. The candidate is shown clearly 
  to be a background object. (In this case, three epochs of data were
  taken to eliminate the possibility that orbital motion of the
  companion would confuse an analysis based on two epochs of data
  alone.)
  \label{fig:uypic}}
%\vskip -0.4in
\end{figure}

A key part of our ongoing observing is to confirm or refute these
candidates via second-epoch measurements, with the first discoveries now
being confirmed ({\bf Figure~\ref{fig:pztel}}).
New companions, especially at the lowest masses, require stringent
validation. Proper motion measurements from second-epoch NICI imaging
will assess if candidates are physically associated with their
primaries. Almost all of our targets have well-measured proper motions
and parallaxes, needed to distinguish background stars from true
companions. The Campaign pipeline delivers high quality astrometry
($\approx$1--5~mas) of very faint (by a factor of $\approx$10$^{5-7}$)
point sources next to bright stars, as validated by fake-companion
injection and multi-epoch measurements of dense stellar fields.
An example of our astrometric performance from high contrast images is
illustrated by the case of UY~Pic ({\bf Figure~\ref{fig:uypic}}), an
AB~Dor moving group member with a very faint 0.8\arcsec\ companion that
we have confirmed to be a background object.

%%%%%%%%%%%%%%%%%%%%%%%%%%%%%%%%%%%%%%%%%%%%%%%%%%%%%%%%%%%%%
\section{Prognosis}

The Campaign is planned to continue for three more semesters, through
the end of 2011. Overall, the Campaign is about halfway done, in terms
of first-epoch observing of the target list. In addition, we are now
transitioning into a new phase, namely the confirmation of new
companions. Astrometric and photometric followup of new discoveries can
be done with NICI alone, as illustrated above. Spectroscopic followup of
the brighter discoveries is feasible with current AO integral-field
spectrographs, as demonstrated by the high-contrast data that have been
obtained with VLT/SINFONI and Keck/OSIRIS (\eg,
\citenum{2007MNRAS.378.1229T, bowler10-hr8799b}). Companions confirmed
at the faintest limits of NICI will be beyond current spectroscopic
capabilities, and analysis will be restricted to photometry alone;
however, such objects will be ideal for detailed studies with
next-generation planet-hunting instruments such as Gemini/GPI and
VLT/SPHERE. In addition, the first-epoch astrometry from NICI
discoveries will be a unique historical resource, which has motivated
our efforts to deliver the highest quality measurements. Determining the
orbital properties of directly imaged planets is a highly desired
long-term goal in this field.

While still in progress, the NICI Planet-Finding Campaign already
represents the largest and most sensitive imaging survey to date for
massive ($\gtrsim$1~\Mjup) planets around other stars. The same Monte
Carlo simulation methods used for the original Campaign planning are
currently being adapted for a uniform analysis of the Campaign
detections (new companions) and non-detections simultaneously, in order
to precisely determine the gas-giant exoplanet mass and separation
distribution. By virtue of its unprecendented sample size, depth, and
uniformity, the Campaign will establish the best measurements to date on
the properties of gas-giant planets at $\gtrsim$5--10~AU separations.

%%%%%%%%%%%%%%%%%%%%%%%%%%%%%%%%%%%%%%%%%%%%%%%%%%%%%%%%%%%%%

% This section describes the normal structure of a manuscript and how each
% part should be handled. The appropriate vertical spacing between various
% parts of this document is achieved in LaTeX through the proper use of
% defined constructs, such as \verb|\section{}|. In LaTeX, paragraphs are
% separated by blank lines in the source file.
%
% At times it may be desired, for formatting reasons, to break a line
% without starting a new paragraph. This situation may occur, for example,
% when formatting the article title, author information, or section
% headings. Line breaks are inserted in LaTeX by entering \verb|\\| or
% \verb|\linebreak| in the LaTeX source file at the desired location.
% 
% %%%%%Sometimes it is necessary to precede the double slash 
% %%%%%by \verb|\protect| to get the desired result, 
% %%%%%for example, in article titles.

%%%%%%%%%%%%%%%%%%%%%%%%%%%%%%%%%%%%%%%%%%%%%%%%%%%%%%%%%%%%%
\acknowledgments     %>>>> equivalent to \section*{ACKNOWLEDGMENTS}       

The NICI Campaign is based on observations obtained at the Gemini
Observatory, which is operated by the Association of Universities for
Research in Astronomy, Inc., under a cooperative agreement with the NSF
on behalf of the Gemini partnership: the National Science Foundation
(United States), the Science and Technology Facilities Council (United
Kingdom), the National Research Council (Canada), CONICYT (Chile), the
Australian Research Council (Australia), Minist\'{e}rio da Ci\^{e}ncia e
Tecnologia (Brazil) and Ministerio de Ciencia, Tecnolog\'{i}a e
Innovaci\'{o}n Productiva (Argentina). Our work was supported in part by
NSF grants AST-0713881 and AST-0709484. B.B. was supported by Hubble
Fellowship grant HST-HF-01204.01-A awarded by the Space Telescope
Science Institute, which is operated by AURA for NASA, under contract
NAS 5-26555.

% This publication makes use of data products from the Two Micron All
% Sky Survey, which is a joint project of the University of
% Massachusetts and the Infrared Processing and Analysis
% Center/California Institute of Technology, funded by the National
% Aeronautics and Space Administration and the National Science
% Foundation.

%%%%%%%%%%%%%%%%%%%%%%%%%%%%%%%%%%%%%%%%%%%%%%%%%%%%%%%%%%%%%
%%%%% References %%%%%

%\bibliography{/Users/mliu/tex/bibtex/mliu.bib}   %>>>> bibliography data in report.bib
%%\bibliography{report}        %>>>> bibliography data in report.bib

\begin{thebibliography}{10}

\bibitem{marois08-hr8799bcd}
Marois, C., {Macintosh}, B., {Barman}, T., {Zuckerman}, B., {Song}, I.,
  {Patience}, J., {Lafreni{\`e}re}, D., and {Doyon}, R., ``{Direct Imaging of
  Multiple Planets Orbiting the Star HR 8799},'' {\em Science}~{\bf 322},  1348
  (Nov. 2008).

\bibitem{2008Sci...322.1345K}
Kalas, P., {Graham}, J.~R., {Chiang}, E., {Fitzgerald}, M.~P., {Clampin}, M.,
  {Kite}, E.~S., {Stapelfeldt}, K., {Marois}, C., and {Krist}, J., ``{Optical
  Images of an Exosolar Planet 25 Light-Years from Earth},'' {\em Science}~{\bf
  322},  1345 (Nov. 2008).

\bibitem{2009A&A...493L..21L}
Lagrange, A. et~al., ``{A Probable Giant Planet Imaged in the {$\beta$}
  Pictoris Disk. VLT/NaCo deep L'-band Imaging},'' {\em \aap}~{\bf 493},
  L21--L25 (Jan. 2009).

\bibitem{2008ApJ...689L.153L}
Lafreni{\`e}re, D., {Jayawardhana}, R., and {van Kerkwijk}, M.~H., ``{Direct
  Imaging and Spectroscopy of a Planetary-Mass Candidate Companion to a Young
  Solar Analog},'' {\em \apjl}~{\bf 689},  L153--L156 (Dec. 2008).

\bibitem{2010ApJ...716..417H}
Hinz, P.~M., {Rodigas}, T.~J., {Kenworthy}, M.~A., {Sivanandam}, S., {Heinze},
  A.~N., {Mamajek}, E.~E., and {Meyer}, M.~R., ``{Thermal Infrared MMTAO
  Observations of the HR~8799 Planetary System},'' {\em \apj}~{\bf 716},
  417--426 (June 2010).

\bibitem{2010arXiv1006.3070L}
Lafreni{\`e}re, D., {Jayawardhana}, R., and {van Kerkwijk}, M.~H., ``{The
  Directly Imaged Planet around the Young Solar Analog 1RXS J160929.1-210524:
  Confirmation of Common Proper Motion, Temperature and Mass},'' {\em ArXiv
  e-prints}  (June 2010).

\bibitem{bowler10-hr8799b}
Bowler, B., Liu, M., Dupuy, T., and Cushing, M., ``{Near-Infrared Spectroscopy
  of the Extrasolar Planet HR~8799~b},'' {\em \apj}  (2010).
\newblock (submitted).

\bibitem{2008SPIE.7015E..49C}
Chun, M. et~al., ``{Performance of the Near-Infrared Coronagraphic Imager on
  Gemini-South},'' in [{\em SPIE Conference
  Series}{\nolinebreak\hspace{0.1em}]},   {\bf 7015} (July 2008).
\newblock arXiv:0809.3017.

\bibitem{1997ApJ...491..856B}
Burrows, A. et~al., ``{A Nongray Theory of Extrasolar Giant Planets and Brown
  Dwarfs},'' {\em \apj}~{\bf 491},  856 (Dec. 1997).

\bibitem{1999PASP..111..587R}
Racine, R., {Walker}, G.~A.~H., {Nadeau}, D., {Doyon}, R., and {Marois}, C.,
  ``{Speckle Noise and the Detection of Faint Companions},'' {\em \pasp}~{\bf
  111},  587--594 (May 1999).

\bibitem{2003IAUS..211..275M}
Marois, C., {Nadeau}, D., {Doyon}, R., {Racine}, R., and {Walker}, G.~A.~H.,
  ``{Differential Simultaneous Imaging and Faint Companions: TRIDENT First
  Results from CFHT},'' in [{\em IAU Symposium}{\nolinebreak\hspace{0.1em}]},
  275 (June 2003).

\bibitem{2004SPIE.5492..970L}
Lenzen, R., {Close}, L., {Brandner}, W., {Biller}, B., and {Hartung}, M., ``{A
  Novel Simultaneous Differential Imager for the Direct Imaging of Giant
  Planets},'' in [{\em Ground-based Instrumentation for Astronomy. Proceedings
  of the SPIE, Volume 5492, pp. 970-977 (2004).}{\nolinebreak\hspace{0.1em}]},
  {Moorwood}, A.~F.~M. and {Iye}, M., eds.,  970--977 (Sept. 2004).

\bibitem{2004SPIE.5490..389B}
Biller, B.~A., {Close}, L., {Lenzen}, R., {Brandner}, W., {McCarthy}, D.~W.,
  {Nielsen}, E., and {Hartung}, M., ``{Suppressing Speckle Noise for
  Simultaneous Differential Extrasolar Planet Imaging (SDI) at the VLT and
  MMT},'' in [{\em Advancements in Adaptive Optics. Edited by Domenico B.
  Calia, Brent L. Ellerbroek, and Roberto Ragazzoni. Proc. of the SPIE, Volume
  5490, pp. 389-397 (2004).}{\nolinebreak\hspace{0.1em}]},   389--397 (Oct.
  2004).

\bibitem{2004Sci...305.1442L}
Liu, M.~C., ``{Substructure in the Circumstellar Disk Around the Young Star AU
  Microscopii},'' {\em Science}~{\bf 305},  1442--1444 (Sept. 2004).

\bibitem{2006ApJ...641..556M}
Marois, C., {Lafreni{\`e}re}, D., {Doyon}, R., {Macintosh}, B., and {Nadeau},
  D., ``{Angular Differential Imaging: A Powerful High-Contrast Imaging
  Technique},'' {\em \apj}~{\bf 641},  556--564 (Apr. 2006).

\bibitem{2010arXiv1005.3084J}
Johnson, J.~A., {Aller}, K.~M., {Howard}, A.~W., and {Crepp}, J.~R., ``{Giant
  Planet Occurrence in the Stellar Mass-Metallicity Plane},'' {\em ArXiv
  e-prints}  (May 2010).

\bibitem{2007ApJ...665..785J}
Johnson, J.~A. et~al., ``{Retired {A} Stars and Their Companions: Exoplanets
  Orbiting Three Intermediate-Mass Subgiants},'' {\em \apj}~{\bf 665},
  785--793 (Aug. 2007).

\bibitem{2008ApJ...674..466N}
Nielsen, E.~L., {Close}, L.~M., {Biller}, B.~A., {Masciadri}, E., and {Lenzen},
  R., ``{Constraints on Extrasolar Planet Populations from VLT NACO/SDI and MMT
  SDI and Direct Adaptive Optics Imaging Surveys: Giant Planets are Rare at
  Large Separations},'' {\em \apj}~{\bf 674},  466--481 (Feb. 2008).

\bibitem{1995MNRAS.273..559J}
Jeffries, R.~D., ``{The Kinematics of Lithium-Rich, Active Late-Type Stars:
  Evidence for a Low-Mass Local Association},'' {\em \mnras}~{\bf 273},
  559--572 (Apr. 1995).

\bibitem{2001ApJ...562L..87Z}
Zuckerman, B., {Song}, I., {Bessell}, M.~S., and {Webb}, R.~A., ``{The
  {$\beta$} Pictoris Moving Group},'' {\em \apjl}~{\bf 562},  L87--L90 (Nov.
  2001).

\bibitem{2008hsf2.book..757T}
Torres, C.~A.~O., {Quast}, G.~R., {Melo}, C.~H.~F., and {Sterzik}, M.~F.,
  [{\em {Young Nearby Loose Associations}}{\nolinebreak\hspace{0.1em}]},  757
  (Dec. 2008).

\bibitem{2009ApJ...699..649S}
Shkolnik, E., {Liu}, M.~C., and {Reid}, I.~N., ``{Identifying the Young
  Low-mass Stars within 25 pc. I. Spectroscopic Observations},'' {\em
  \apj}~{\bf 699},  649--666 (July 2009).

\bibitem{2008AJ....136.1290R}
Reid, I.~N. et~al., ``{Meeting the Cool Neighbors. X. Ultracool Dwarfs from the
  2MASS All-Sky Data Release},'' {\em \aj}~{\bf 136},  1290--1311 (Sept. 2008).

\bibitem{2010ApJ...717..878N}
Nielsen, E.~L. and {Close}, L.~M., ``{A Uniform Analysis of 118 Stars with
  High-contrast Imaging: Long-period Extrasolar Giant Planets are Rare Around
  Sun-like Stars},'' {\em \apj}~{\bf 717},  878--896 (July 2010).

\bibitem{2008SPIE.7015E.184B}
Biller, B. et~al., ``{Observing Strategies for the NICI Campaign to Directly
  Image Extrasolar Planets},'' in [{\em SPIE Conference
  Series}{\nolinebreak\hspace{0.1em}]},   {\bf 7015} (July 2008).
\newblock arXiv:0809.3020.

\bibitem{2007ApJ...670.1367L}
Lafreni{\`e}re, D. et~al., ``{The Gemini Deep Planet Survey},'' {\em \apj}~{\bf
  670},  1367--1390 (Dec. 2007).

\bibitem{2010ApJ...716.1551L}
Leconte, J. et~al., ``{The Lyot Project Direct Imaging Survey of Substellar
  Companions: Statistical Analysis and Information from Nondetections},'' {\em
  \apj}~{\bf 716},  1551--1565 (June 2010).

\bibitem{2007ApJS..173..143B}
Biller, B.~A. et~al., ``{An Imaging Survey for Extrasolar Planets around 45
  Close, Young Stars with the Simultaneous Differential Imager at the Very
  Large Telescope and MMT},'' {\em \apjs}~{\bf 173},  143--165 (Nov. 2007).

\bibitem{1980ApJS...44...73B}
Bahcall, J.~N. and {Soneira}, R.~M., ``{The universe at faint magnitudes. I -
  Models for the galaxy and the predicted star counts},'' {\em \apjs}~{\bf 44},
   73--110 (Sept. 1980).

\bibitem{akraus2009-thesis}
Kraus, A.~L., {\em {Multiple Star Formation}}, PhD thesis, Caltech (2009).

\bibitem{2000ApJ...542..464C}
Chabrier, G., {Baraffe}, I., {Allard}, F., and {Hauschildt}, P.,
  ``{Evolutionary Models for Very Low-Mass Stars and Brown Dwarfs with Dusty
  Atmospheres},'' {\em \apj}~{\bf 542},  464--472 (Oct. 2000).

\bibitem{biller10-PZTelB}
Biller, B.~A. et~al., ``{The Gemini NICI Planet-Finding Campaign: Discovery of
  a Close Substellar Companion to the Nearby Young Solar Analog PZ~Tel},'' {\em
  \apjl}  (2010).
\newblock in press.

\bibitem{2007MNRAS.378.1229T}
Thatte, N., {Abuter}, R., {Tecza}, M., {Nielsen}, E.~L., {Clarke}, F.~J., and
  {Close}, L.~M., ``{Very High Contrast Integral Field Spectroscopy of AB
  Doradus C: 9-mag Contrast at 0.2\arcsec\ Without a Coronagraph Using Spectral
  Deconvolution},'' {\em \mnras}~{\bf 378},  1229--1236 (July 2007).

\end{thebibliography}
%\bibliographystyle{spiebib}   %>>>> makes bibtex use spiebib.bst

\end{document}